\documentstyle[11pt]{article}
\begin{document}
\begin{center}
{\large\bf
New Regularization-Renormalization
Method in Quantum Electrodynamics and Qualitative Calculation on Lamb Shift}
\end{center}
\vspace{.7cm}
\centerline{Guang-jiong Ni$^*$\footnotetext{$^*$E-mail:
gjni@fudan.ihep.ac.cn} and Haibin Wang}
\vspace{0.2cm}
\centerline{Department of Physics,
Fudan University, Shanghai 200433, China}
\vspace{0.7cm}
\centerline{abstract}
A simple but effective method for regularization-renormalization (R-R)
is proposed for handling the Feynman diagram integral (FDI) at one loop
level in quantum electrodynamics (QED). The divergence is substituted by
some constants to be fixed via experiments. So no counter term, no bare
parameter and no arbitrary running mass scale is involved. Then the Lamb
Shift in Hydrogen atom can be calculated qualitatively and simply as
$\Delta E(2S_{1/2})- \Delta E(2P_{1/2})=996.7 MHz$ versus the experimental
value $1057.85 MHz$.
\vspace*{5mm}

 
\newpage
In quantum field theory (QFT), when calculating the Feynman diagram
integral (FDI) beyond the tree level, one soon encountered the
divergence. After handling it by some regularization method, one further
introduced counter terms for cancelling the divergence and established
the relation between bare parameter and physical parameter. Many
physicists are not satisfyed with the above recipe. Can the divergence be
avoided? What happens when one pushes the cutoff $\Lambda$ in momentum
integral into infinity? Is this implying the point-like feature in QFT
model? Will we need a drastic change in the foundation of our theory?
(see the final discussion in the excellent book by Sakurai [1].)

After learning the previous methods for regularization-renormalization
(R-R)[2], begining from the work by Yang and Ni [3] and further by Ni
et al. [4,5], we proposed a new R-R method as follows. When encountering
a superficially divergent FDI, we first differentiate it with respect to
external momentum or mass parameter enough times until it becomes
convergent. After performing integration with respect to internal
momentum, we reintegrate it with respect to the parameter the same times
to return to original FDI. Then instead of divergence, some
arbitrary constants $C_{i}$($i=1,2,\cdots$) appear in FDI, showing the
lack of knowledge about the model at QFT level under consideration. So
they should be fixed by experiment via suitable renormalization
procedure. For illustration, let us consider the various FDI at one loop
level in QED, using Bjorken-Drell metric [6].

1. Self-energy of electron in QED.

The FDI for self-energy of
electron reads (see [1,6,7,8]) ($e<0$)
\begin{eqnarray}
-i\Sigma(p) &=& (-ie)^{2}\int\frac{d^{4}k}{(2\pi)^{4}}
\frac{g_{\mu\nu}}{ik^{2}}\gamma^{\mu}\frac{i}{\not{p}-
\not{k~~}-m}
\gamma^{\nu} \nonumber \\
&=& -e^{2}\int\frac{d^{4}k}{(2\pi)^{4}}\frac{N}{D}
\end{eqnarray}
$$
\frac{1}{D}=\frac{1}{k^{2}[(p-k)^{2}-m^{2}]}
=\int_{0}^{1}\frac{dx}{[k^{2}-2p\cdot kx+(p^{2}-m^{2})x]^{2}}
$$
$$
N=g_{\mu\nu}\gamma^{\mu}(\not{p}-\not{k}+m)\gamma^{\nu}
=-2(\not{p}-\not{k})+4m
$$
We first perform a shift in momentum integration: $k\rightarrow K=k-xp$,
so that
\begin{equation}
-i\Sigma(p)=-e^{2}\int_{0}^{1}dx[-2(1-x)\not{p}+4m]I
\end{equation}
and concentrate on the logarithmically divergent integral:
\begin{equation}
I=\int\frac{d^{4}K}{(2\pi)^{4}}\frac{1}{[K^{2}-M^{2}]^{2}}
\end{equation}
with
\begin{equation}
M^{2}=p^{2}x^{2}+(m^{2}-p^{2})x
\end{equation}
A differentiation with respect to $M^2$ is enough to get
\begin{equation}
\frac{\partial{I}}{\partial M^{2}}=\frac{-i}{(4\pi)^{2}}\frac{1}{M^{2}}
\end{equation}
(see formula (C.2.5) in Ref.[9]). Thus
\begin{equation}
I=\frac{-i}{(4\pi)^{2}}[\ln M^{2}+C_{1}]=\frac{-i}{(4\pi)^{2}}\ln
\frac{M^{2}}{\mu^{2}_{2}}
\end{equation}
carries an arbitrary contant $C_{1}=-\ln\mu_{2}^{2}$. After integration
with respect to the Feynman parameter $x$, one obtains ($\alpha\equiv
e^{2}/4\pi$)
\begin{eqnarray}
\Sigma(p)&=& A+B\not{p} \nonumber \\
A &=& \frac{\alpha}{\pi}m[2-\ln
\frac{m^{2}}{\mu^{2}_{2}}+\frac{(m^{2}-p^{2})}{p^{2}}\ln
\frac{(m^{2}-p^{2})}{m^{2}}] \nonumber \\
B &=& \frac{\alpha}{4\pi}\{\ln
\frac{m^{2}}{\mu^{2}_{2}}-3-\frac{(m^{2}-p^{2})}{p^{2}}
[1+\frac{m^{2}+p^{2}}{p^{2}}\ln\frac{(m^{2}-p^{2})}{m^{2}}]\}
\end{eqnarray}
Using the chain approximation, one can derive the modification of
electron propagator as:
\begin{equation}
\frac{i}{\not{p}-m}\rightarrow \frac{i}{\not{p}-m}\frac{1}{1
-\frac{\Sigma(p)}{\not{p}-m}}=
\frac{iZ_{2}}{\not{p}-m_{\rm R}}
\end{equation}
\begin{equation}
Z_{2}=(1-B)^{-1}\simeq 1+B
\end{equation}
\begin{equation}
m_{\rm R}=\frac{m+A}{1-B}\simeq(m+A)(1+B)\simeq m+\delta m
\end{equation}
$$
\delta m\simeq A+mB
$$
For a free electron, the mass shell condition $p^{2}=m^{2}$ leads to
$$
\delta m=\frac{\alpha m}{4\pi}(5-3\ln\frac{m^{2}}{\mu_{2}^{2}})
$$
We want the parameter $m$ in the Lagrangian still being explained as the
observed mass, i.e., $m_{\rm R}=m_{\rm obs}=m$. So $\delta {m}=0$ leads
to $\ln\frac{m^2}{\mu_{2}^{2}}=\frac{5}{3}$, which in turn fixes the
renormalization factor for wave function
\begin{equation}
Z_{2}=1-\frac{\alpha}{3\pi}
\end{equation}

2. Photon self-energy---vacuum polarization.
\begin{equation}
\Pi_{\mu\nu}(q)=-(-ie)^{2}{\rm Tr}\int\frac{d^{4}k}{(2\pi)^{4}}
\gamma_{\mu}\frac{i}{\not{k}-m}\gamma_{\nu}\frac{i}{\not{k}-\not{q}-m}
\end{equation}
Introducing the Feynman parameter $x$ as before and performing a shift
in momentum integration: $k\rightarrow K=k-xq$, we get
\begin{equation}
\Pi_{\mu\nu}(q)=-4e^{2}\int_{0}^{1}dx(I_{1}+I_{2})
\end{equation}
where
\begin{equation}
I_{1}=\int\frac{d^{4}K}{(2\pi)^{4}}\frac{2K_{\mu}K_{\nu}-g_{\mu\nu}K^{2}}
{(K^{2}-M^{2})^{2}}
\end{equation}
with
\begin{equation}
M^{2}=m^{2}+q^{2}(x^{2}-x)
\end{equation}
is quadratically divergent while
\begin{equation}
I_{2}=\int\frac{d^{4}K}{(2\pi)^{4}}\frac{(x^{2}-x)(2q_{\mu}q_{\nu}-
g_{\mu\nu}q^{2})+m^{2}g_{\mu\nu}}{(K^{2}-M^{2})^{2}}
\end{equation}
is only logarithmically divergent like that in Eqs.(3-6).
An elegant way for handling $I_1$ is modifying $M^2$ into
\begin{equation}
M^{2}(\sigma)=m^{2}+q^{2}(x^{2}-x)+\sigma
\end{equation}
and differentiating $I_1$ with respect to $\sigma$
two times. After integration
with respect to $K$, we reintegrate it with respect to $\sigma$ two
times, arriving at the limit $\sigma\rightarrow 0$:
\begin{equation}
I_{1}=\frac{ig_{\mu\nu}}{(4\pi)^{2}}\{[m^{2}+q^{2}(x^{2}-x)]
\ln\frac{m^{2}+q^{2}(x^{2}-x)}{\mu_{3}^{2}}+C_{2}\}
\end{equation}
with two arbitrary constants: $C_{1}=-\ln\mu_{3}^{2}$ and $C_{2}$.
Combining $I_{1}$ and $I_2$ together, we find
\begin{equation}
\Pi_{\mu\nu}(q)=\frac{8ie^{2}}{(4\pi)^{2}}(q_{\mu}q_{\nu}-g_{\mu\nu}q^{2})
\int_{0}^{1}dx(x^{2}-x)\ln\frac{m^{2}+q^{2}(x^{2}-x)}{\mu_{3}^{2}}
-\frac{i4e^{2}}{(4\pi)^{2}}g_{\mu\nu}C_{2}
\end{equation}
The continuity equation of current induced in the vacuum polarization
[1]
\begin{equation}
q^{\mu}\Pi_{\mu\nu}(q)=0
\end{equation}
is ensured by the factor ($q_{\mu}q_{\nu}-g_{\mu\nu}q^2$). So we set
$C_{2}=0$. Consider the scattering between two electrons via the
exchange of a photon with momentum transfer $q\rightarrow 0$ [1].
Adding the contribution of $\Pi_{\mu\nu}(q)$ to tree diagram amounts to
modify the charge square:
$$
e^{2}\rightarrow e_{\rm R}^{2}=Z_{3}e^{2}
$$
\begin{equation}
Z_{3}=1+\frac{\alpha}{3\pi}(\ln\frac{m^{2}}{\mu_{3}^{2}}-
\frac{q^{2}}{5m^{2}}+\cdots)
\end{equation}
The choice of $\mu_{3}$ will be discussed later. The next term in
expansion when $q\neq 0$ constributes a modification on Coulumb
potential due to vacuum polarization (Uehling potential).

3. Vertex function in QED.
\begin{equation}
\Lambda_{\mu}(p',p)=(-ie)^{2}\int\frac{d^{4}k}{(2\pi)^{4}}\frac{-i}{k^{2}}
\gamma_{\nu}\frac{i}{\not{p'}-\not{k}-m}\gamma_{\mu}\frac{i}{\not{p}-\not{k}-m}
\gamma^{\nu}
\end{equation}
For simplicity, we consider electron being on the mass shell:
$p^{2}={p'}^{2}=m^2$, $p'-p=q$, $p\cdot q=-\frac{q^{2}}{2}$. Introducing the
Feynman parameter $u=x+y$ and $v=x-y$, we perform a shift in momentum
integration:
$$
k\rightarrow K=k-(p+\frac{q}{2})u-\frac{q}{2}v
$$
Thus
\begin{equation}
\Lambda_{\mu}=-ie^{2}[I_{3}
\gamma_{\mu}+I_{4}]
\end{equation}
\begin{equation}
I_{3}=\int_{0}^{1}du\int_{-u}^{u}dv\int\frac{d^{4}K}{(2\pi)^{4}}
\frac{K^{2}}{(K^{2}-M^{2})^{3}}
\end{equation}
\begin{equation}
M^{2}=(m^{2}-\frac{q^{2}}{4})u^{2}+\frac{q^{2}}{4}v^{2}
\end{equation}
\begin{equation}
I_{4}=\int_{0}^{1}du\int_{-u}^{u}dv\int\frac{d^{4}K}{(2\pi)^{4}}
\frac{A_{\mu}}{(K^{2}-M^{2})^{3}}
\end{equation}
\begin{eqnarray}
A_{\mu}&=&(4-4u-2u^{2})m^{2}\gamma_{\mu}+2i(u^{2}-u)mq^{\nu}\sigma_{\mu\nu}
\nonumber \\
& & -(2-2u+\frac{u^{2}}{2}-\frac{v^{2}}{2})q^{2}\gamma_{\mu}
-(2+2u)vmq_{\mu}
\end{eqnarray}
Set $K^{2}=K^{2}-M^{2}+M^{2}$,then $I_{3}=I_{3}^{'}-\frac{i}{32\Pi^{2}}$.
$I_{3}^{'}$ is only logarithmically divergent and can be treated as before
to be
\begin{equation}
I_{3}=
\frac{-i}{(4\pi)^{2}}\int_{0}^{1}du\int_{-u}^{u}dv
\ln\frac{(m^{2}-\frac{q^{2}}{4})u^{2}+\frac{q^{2}}{4}v^{2}}{\mu_{1}^{2}}
\end{equation}
with $\mu_{1}^{2}$ an arbitrary constant. Now $q^{2}=-Q^{2}<0$
($Q^{2}>0$)
\begin{equation}
I_{3}=\frac{-i}{(4\pi)^{2}}(\ln\frac{m^{2}}{\mu_{1}^{2}}
-\frac{5}{2}+\frac{1}{w}F(w)\}
\end{equation}
$$
F(w)=\ln \frac{1+w}{1-w}
$$
$$
w=\frac{1}{\sqrt{\frac{4m^2}{Q^2}+1}}
$$
On the other hand, though there is no ultra-violet divergence in
$I_{4}$, it does have infrared divergence at $u \rightarrow 0$. 
For handling it, we
introduce  a lower cutoff $\eta$ in the integration with respect
to $u$
\begin{eqnarray}
I_{4}&=&\frac{i}{2(4\pi)^{2}}\{[4\ln\eta+5]\frac{4w}{Q^2}F(w)m^{2}
\gamma_{\mu}+\frac{i4w}{Q^{2}}F(w)mq^{\nu}\sigma_{\mu\nu}  \nonumber \\
& &
+4(2\ln\eta+\frac{7}{4})wF(w)\gamma_{\mu}+[\frac{1}{w}F(w)-2]\gamma_{\mu}\}
\end{eqnarray}
Combining Eqs.(29) and (30) into Eq.(23), one arrives at
\begin{eqnarray}
\Lambda_{\mu}(p',p)&=&
-\frac{\alpha}{4\pi}\{[\ln\frac{m^{2}}{\mu_{1}^{2}}-\frac{3}{2}+\frac{1}{2w}F(w)]
\gamma_{\mu}-(4\ln\eta+5)\frac{2w}{Q^2}F(w)m^{2}\gamma_{\mu}
\nonumber \\
& & -\frac{i2w}{Q^2}F(w)mq^{\nu}\sigma_{\mu\nu}-2(2\ln\eta+\frac{7}{4})
wF(w)\gamma_{\mu}\}
\end{eqnarray}
When ($Q^{2}<<m^{2}$), we get
$$
\Lambda_{\mu}(p',p)=\frac{\alpha}{4\pi}(\frac{11}{2}-\ln\frac{m^{2}}{\mu_{1}^{2}}
+4\ln\eta)\gamma_{\mu}+i\frac{\alpha}{4\pi}
\frac{q^{\nu}}{m}\sigma_{\mu\nu}-\frac{\alpha}{4\pi}(
\frac{1}{6}+\frac{4}{3}\ln\eta)\frac{q^{2}}{m^{2}}\gamma_{\mu}
$$
It means the interaction of the electron with the external potential is
modified
\begin{equation}
-e\gamma_{\mu}\rightarrow -e[\gamma_{\mu}+\Lambda_{\mu}(p',p)]
\end{equation}
Besides the important term
$i\frac{\alpha}{4\pi}\frac{q^{\nu}}{m}\sigma_{\mu\nu}$ which emerges as
the anomalous magnetic moment of electron, the charge modification here
is expressed by a renormalization factor $Z_1$:
\begin{equation}
Z_{1}^{-1}=1+\frac{\alpha}{4\pi}\{[2-\ln\frac{m^2}{\mu_{1}^2}-\frac{1}{2w}F(w)]
+(4\ln\eta+5)\frac{2w m^{2}}{Q^{2}}F(w)+(2\ln\eta+\frac{7}{4})2wF(w)\}
\end{equation}
The infrared term ($\sim\ln\eta$) is ascribed to
the bremsstrahlung of soft photons [6,8] and can be taken care by KLN
theorem [10]. We will fix $\mu_{1}$ and $\eta$ below.

4. Beta function at one loop level.

Adding all four FDI's at one loop level to the tree diagram, we define
the renormalized charge as usual[6-9]:
\begin{equation}
e_{\rm R}=\frac{Z_{2}}{Z_{1}}Z_{3}^{1/2}e
\end{equation}
But the Ward-Takahashi Identity (WTI) implies that [6-8]
\begin{equation}
Z_{1}=Z_{2}
\end{equation}
Therefore
\begin{equation}
\alpha_{\rm R}\equiv\frac{e_{\rm R}^{2}}{4\pi}=Z_{3}\alpha
\end{equation}
Then set $p^{2}=m^{2}$ in $Z_{2}$ and $Q^{2}=0$ in $Z_{1}$ with $\mu_{1}=
\mu_{2}$, yielding
\begin{equation}
\ln\eta=-\frac{5}{8}
\end{equation}
For any value of $Q$, the renormalized charge reads
from Eqs. (19-21):
\begin{equation}
e_{\rm R}(Q)=e\{1+\frac{\alpha}{\pi}\int_{0}^{1}dx[(x-x^{2})\ln
\frac{Q^{2}(x-x^{2})+m^{2}}{\mu_{3}^{2}}]\}
\end{equation}
\begin{equation}
e_{\rm R}(Q)\sim
e\{1+\frac{\alpha}{2\pi}[\frac{1}{3}\ln\frac{m^{2}}{\mu_{3}^{2}}+
\frac{1}{15}\frac{Q^{2}}{m^{2}}]\},~~ (Q^{2}<<m^{2})
\end{equation}
The observed charge is defined at $Q^{2}\rightarrow 0$ (Thomson
scattering) limit:
\begin{equation}
e_{\rm obs}=e_{\rm R}|_{Q=0}=e
\end{equation}
which dictates that
\begin{equation}
\mu_{3}=m
\end{equation}
We see that $e^{2}_{R}(Q)$ increases with $Q^{2}$. For
discussing the running of $\alpha_{\rm R}$ with $Q^{2}$, we define the
Beta function:
\begin{equation}
\beta(\alpha, Q)\equiv Q\frac{\partial}{\partial Q}\alpha_{\rm R}(Q)
\end{equation}
From Eq.(38), one finds:
\begin{equation}
\beta(\alpha,Q)=\frac{2\alpha^{2}}{3\pi}-
\frac{4\alpha^{2}m^{2}}{\pi
Q^{2}}\{1+\frac{2m^{2}}{\sqrt{Q^{4}+4Q^{2}m^{2}}}
\ln\frac{\sqrt{Q^{4}+4Q^{2}m^{2}}-Q^{2}}
{\sqrt{Q^{4}+4Q^{2}m^{2}}+Q^{2}}\}
\end{equation}
\begin{equation}
\beta(\alpha, Q)\simeq\frac{2\alpha^{2}}{15\pi}
\frac{Q^{2}}{m^{2}},~~ (\frac{Q^{2}}{4m^{2}}<<1)
\end{equation}
\begin{equation}
\beta(\alpha, Q)\simeq\frac{2\alpha^{2}}{3\pi}-
\frac{4\alpha^{2}m^{2}}{\pi Q^{2}},~~
(\frac{4m^{2}}{Q^{2}}<<1)
\end{equation}
which leads to the well known result $\beta(\alpha)=\frac{2\alpha^{2}}
{3\pi}$ at one loop level at $Q^{2} \rightarrow \infty$.

5. The renormalization group equation

The renormalization group equation (RGE) in QED is ontained by setting
$Q \rightarrow \infty$ and $\alpha \rightarrow \alpha_{R}(Q)$ in the right
hand side of $Eq. (42)$,
\begin{equation}
Q\frac{\partial}{\partial Q}\alpha_{R}=\frac{2 \alpha^{2}_{R}}{3 \pi}
\end{equation}
Then after integration, one yields
\begin{equation} 
\alpha_{R}(Q)=\frac{\alpha}{1-\frac{2 \alpha}{3 \pi}\ln\frac{Q}{m}}
\end{equation}
where the renormalization is made as
\begin{equation}
\alpha_{R}|_{Q=m}=\alpha
\end{equation}
Note that, however, the running physical parameter is momentum transfer $Q$,
not $\mu$. All $\mu_{i}(i=1,2,3)$
had been fixed.

A Landau pole emerges at high $Q$,
\begin{equation}
Q_{landau}\equiv\mu_{Landau}=m exp(\frac{3 \pi}{2 \alpha})
\end{equation}

6. The caculation of Lamb Shift.

The famous Lamb Shift in Hydrogen atom, $i.e.$, the energy difference
between 2S$_{1/2}$ and 2P$_{1/2}$ states[11],
\begin{equation}
[\Delta E(2 {\rm S}_{1/2})-\Delta E(2 {\rm P}_{1/2})]_{exp}=1057.845(9) MHz
\end{equation}

can be calculated in our formalism qualitatively in an elegant manner.

In $Eqs. (7)-(10)$, we set
\begin{equation}
\delta m=A+mB=-\frac{\alpha^{2} m}{2 n^{2}}
\end{equation}
and a deviation parameter $\eta$ from the mass shell of free motion,
\begin{equation}
p^{2}=m^{2}(1-\eta)
\end{equation}
This parameter is precisely the same $\eta$ used in $Eqs. (30)-(33)$ for
avoiding the infrared divergence in vertex function. Then
\begin{equation}
-\frac{\alpha^{2} m}{2 n^{2}}=\frac{\alpha m}{4 \pi}
  \frac{(-\eta + 2\eta \ln\eta)}{1+\frac{\alpha}{3 \pi}}
\end{equation}
For $n=2$ case,
\begin{eqnarray}
\eta=\eta_{2}=7.44489 \times 10^{-4} \nonumber \\
\ln\eta_{2}=-7.20281
\end{eqnarray}

Now the $q^{2}$ term in vertex function implies a modification on Coulomb
potential,
\begin{equation}
-\frac{e^{2}}{4 \pi r} \rightarrow -\frac{e^{2}}{4 \pi r}
    - \frac{\alpha^{2}}{m^{2}}(\frac{1}{6}+\frac{4}{3}\ln\eta)\delta(\vec{r})
\end{equation}
Adding further the contibution from anomalous magnetic momentum in vertex
function and the Uehling potential in $Eq. (21)$, one finds the energy shift
of nS$_{1/2}$ state as,
\begin{equation}
\Delta E(nS_{1/2})= \frac{\alpha^{3}}{3 \pi n^{3}} R_{y} (\frac{1}{20}
+\ln\frac{1}{\eta})
\end{equation}
where $R_{y}\equiv \frac{1}{2}\alpha^{2} m$.

Then
\begin{equation}
\Delta E(2S_{1/2})= 979.73 MHz 
\end{equation}
\begin{equation}
[ \Delta E(2S_{1/2}) - \Delta E(2P_{1/2})]_{Theor} = 996.69 MHz
\end{equation}

where 2P$_{1/2}$ state gets an extra downward shift due to the anomalous
magnetic moment.

In recent years, the 'absolute' Lamb Shift of 1S$_{1/2}$ state was measured
as
\begin{equation}
[\Delta E(1S_{1/2})]_{exp}=8172.86(5) MHz
\end{equation}
In the above calculation, we get
\begin{equation}
[\Delta E(1S_{1/2})]_{Theor}=6076.79 MHz
\end{equation}
with
\begin{eqnarray}
\eta=\eta_{1}=3.77274 \times 10^{-3}    \\
\ln\eta_{1}=-5.57995
\end{eqnarray}
But we should add extra contributions from the vaccum polarization with
$\mu_{3}=m$, but $m \rightarrow m-R_{y}/n^{2}$ 
\begin{equation}
[\Delta E^{VP}(1S_{1/2})]=\frac{2 Z^{4} \alpha^{3}}{3 \pi n^{4}} R_{y}
    |_{Z=1, n=1}=271 MHz
\end{equation}
and the finite radius of proton $r_{p}=0.862 \times 10^{-13} cm$,
\begin{equation}
[\Delta E^{r_{p}}(1S_{1/2})]=\frac{4}{5}R_{y}(\frac{r_{p}}{a_{0}})^{2}
   =0.7 MHz
\end{equation}
Altogether, we obtain
\begin{equation}
[\Delta E(1S_{1/2})]_{Theor} \simeq 6349 MHz
\end{equation}

In summary, some remarks are in order.

(a) Our R-R method is really very simple: (i) When encountering a
superficially divergent FDI, we use Feynman parameter trick to combine
the donominator D into one factor. (ii) Perform a momentum shift from
$k\rightarrow K$, so that $D\sim (K^{2}-M^{2})^{n}$; (iii) Take derivative
of FDI with respect to $M^{2}$ to raise the value of $n$ such that the
integral becomes convergent. (iv) After momentum integration,
reintegrate it with respect to $M^{2}$. (v) Then some arbitrary
constants ($\mu_{i}, C_{2}$) with $\eta$
emerge. They can only be fixed by
experiments (the observed mass $m$ and charge $e$) or by some deep reason in
theoretical consideration (like the continuity condition of current or
WTI). (vi) Since all constants $(\mu_{\rm i}, C_{\rm 2}, \eta)$
are fixed at one loop level (L=1) with the meaning of $m$ and $e$
reconfirmed as that at tree level, all previous steps can be repeated at
next loop expansion (L=2).

(b) In QED, like any renormalizable model in QFT, there will be no
trouble in high L calculations because the number of arbitrary constants
corresponds to that of so-called primitive divergent integrals, whose
number remains finite. On the other hand, in a nonrenormalizable model,
there would be more and more arbitrary constants $C_{i}$'s in high L
calculation, showing that such kind of model is not well-defined at
QFT level.

(c) The procedure of shifting momentum $k \rightarrow K$ is legal in our
treatment because, (i) For a logrithmically divergent FDI, the momentum
shift brings no change in value; (ii) For a linearly or quadradically
divergent FDI, the shift does bring a change (surface integal term) in
the value, but eventually it is absorbed into arbitrary constants.

(d) The reason why these constants appear in substitution of divergence
was explained in detail in Ref.[4]. It reflects the fact that the
world is infinite whereas our knowledge remains finite. In particular,
the value of mass $m$ or charge $e$ is beyond the reach of perturbative
QED. The calculation on self-energy $\Sigma$ as in this paper has nothing
to do with the mass generation of electron except a finite and fixed
renormalization on wave function: $Z_{2}=1-\frac{\alpha}{3\pi}$.
A model how a fermion can acquire a mass together with the phase
transition of vacuum (environment) which provides a second mass scale (the
standard weight) was discussed in Ref.[12]. Besides, mass of electron
can be modified via radiative correction on self-energy when it is
moving inside a cavity (see Ref.[13]). The appearence of Landau pole
is also inevitable, showing  the boundary of QED theory(see [5]).

(e) No counter term and/or bare parameter is needed any more.

(f) There is also not any arbitrary  running mass scale after
renormalizatiion. All arbitrary conatants $\mu_{i}$and $\eta$ should be
fixed in renormalization, leaving only the physical running mass scale
like external momentum $p$ or momentum transfer $Q$.

(g) It is interesting to see the qualitative calculation of Lamb Shift
being so simple. The discrepency between theoretical and experimental
values reflects the fact that the covariant form of QFT is not suitable
for dealing with the binding state problem.

(h) This R-R method was used to derive the Higgs mass,
138 GeV, in standard model [4-5]. It works also quite well in QCD, which will
be discussed elsewhere.

We thank Profs. S-q Chen, Y-s Duan, B-y Hou, T. Huang, J-w Qiu, K-j Shi, B.L.
Young, Z-x Zhang, Z-y Zhu
and Drs. J-f Yang and Z-g Yu for discussions.
This work
was supported in part by the National Science Foundation in China.

\vspace*{2cm}
\centerline{References}

\noindent [1] J. J. Sakurai, {\it Advanced Quantum Mechanics},
Addison-Wesley Publishing Company, 1967.

\noindent [2] H. Epstein and V. Glaser, Ann. Inst. Hemi. Poincare 19,
211(1973); G. Scharf, Finite Electrodynamics, Spring-Verlag, Berlin,1989;
J.Glimm and A. Jaffe, Collective Papers, Vol. 2(1985);
J. D$\ddot{u}$tch, F. Krahe and G.scharf, Phys. Lett. B258, 457(1991);
D. Z. Freedmann, K. Johnson and J. I. Lattore, Nucl. Phys. B371, 353(1992).

\noindent [3] Ji-feng Yang, Thesis for PhD degree, 1994, unpublished.
J-f Yang and G-j Ni, Acta Physica Sinica {\bf 4}, 88(1995).

\noindent [4] G-j Ni and S-q Chen, preprint.

\noindent [5] G-j Ni, S-q Chen, S-y Lou, W-f Lu and J-f Yang, Proceeding
of the 7th Conference on Particle Physics in China, 1996, p55-59. G-j Ni,
S-y Lou, W-f Lu and J-f Yang, preprint.

\noindent [6] J. D. Bjorken and S. D. Drell, {\it Relativistic Quantum
Mechanics}, McGraw-Hill Book Company, 1964.

\noindent [7] P. Ramond, {\it Field Theory: A Modern Primer}, The
Benjamin/Cummings Publishing Company, 1981.

\noindent [8] C. Itzykson and J. B. Zuber, {\it Qunatum Field Theory},
McGraw-Hill Inc., 1980.

\noindent [9] R. D. Field, {\it Application of Perturbative QCD},
Addison-Wesley Publishing Company, 1989.

\noindent [10] T. D. Lee, {\it Particle Physics and Introduction to
Field Theory}, Harwood Academic publishers, 1981.

\noindent [11] S. Weinberg, The Quantum Theory of Fields, I. Foundations
(Cambridge University Press) 1995, p.578-596.

\noindent [12] G-j Ni, J-f Yang, D-h Xu and S-q Chen, Commun. Theor.
Phys. {\bf 21}, 73 (1994).

\noindent [13] G-j Ni and S-q Chen, Chin. Phys. Lett., 14,9(1997).

\end{document}